\documentclass[12pt,a4paper,onecolumn]{article}

\usepackage[top=2cm,bottom=2cm,left=2cm,right=2cm]{geometry} 
\usepackage{enumerate}
\usepackage{graphicx}
\usepackage{caption}
\usepackage{xcolor}
\usepackage[normalem]{ulem}

\title{Multiple folding pathways of proteins with shallow knots  and co-translational folding}

\author{Mateusz Chwastyk and Marek Cieplak\\
\small Institute of Physics, Polish Academy of Sciences, Al. Lotnik\'ow 32/46, 02-668 Warsaw, Poland}

\begin{document}

\maketitle

\abstract{
We study the folding process in the shallowly knotted protein MJ0366
within two variants of a structure-based model. 
We observe that the resulting topological pathways are much richer
than identified in previous studies.
In addition to the single knot-loop events,
we find novel, and dominant, two-loop mechanisms. 
We demonstrate that folding takes place in a range of temperatures
and the conditions of most successful folding are at temperatures
which are higher than those required for the fastest folding. 
We also demonstrate that nascent conditions are more favorable
to knotting than off-ribosome folding. 
}

\section{Introduction}

Sufficiently long polymers, such as DNA, are 
likely to be entangled \cite{DNA1,DNA}.  On the other hand, there are very 
few knotted RNA molecules -- in a recent assessment \cite{RNA} only three
cases have been identified, but the corresponding structures
are resolved poorly.
Occurrence of knotted proteins is in between
-- several hundreds of knot-containing structure files
and a dozen of truly independent structures
\cite{Virnau,Virnau1,Stasiak,KnotProt} in the Protein Data Bank (PDB).
Knots in proteins can be detected experimentally through stretching \cite{Rief}
since their effective contour length is reduced, as analyzed theoretically
in refs. \cite{Sulkowska_2008,JACS,Dziubiella,Israel}. However, 
figuring out how knots get formed during the folding process is
more challenging. New experimental techniques based on fusion proteins
\cite{fusion,LimJackson} have started to offer clues about the process.
Nevertheless computer simulations are expected to offer more
detailed insights. In this paper,
we analyze pathways of folding in a model shallowly knotted protein
and reveal a rich complexity of possible behaviors. We also propose 
a mechanical model of the ribosome and show that nascent conditions
favor knotting but also affect the pathways of folding.

Backbones of proteins do not form closed loops which leads to
some ambiguities when  deciding about the presence of a knot.
Nevertheless, it is often straightforward to identify knot ends by 
observing knot's unknotting 
on cutting away sites from the termini
\cite{km1,taylor}. A knot is considered shallow if at least one of
its ends is close to a terminus. Otherwise,
the knot is considered to be deep.

Deeply knotted proteins such as YibK
have been studied experimentally \cite{Mallam,Mallam1}.
They are known to fold with difficulty in simulations.
A folding process here is considered successful
if the native contacts are established \emph{and} the knot
is formed properly.
The success rate, $S$, is, at best, 1 - 2 \% \cite{dodging,Takada}.
We have recently argued \cite{nascent} that nascent 
conditions \cite{Dobson,Bustamante,ribosome,ribosome1,ribosome2}
enhance the probability for YibK to become knotted when formed on the
ribosome. We have confirmed
that the successful folding pathway goes through a slipknot conformation,
as suggested in ref. \cite{dodging} for the off-ribosome situation. We have
shown that the process takes place only in a range of optimal temperatures,
and does not require any non-native interactions (if a proper procedure for 
their selection is adopted).

\begin{figure}[ht]
\begin{center}
\includegraphics[scale=0.35]{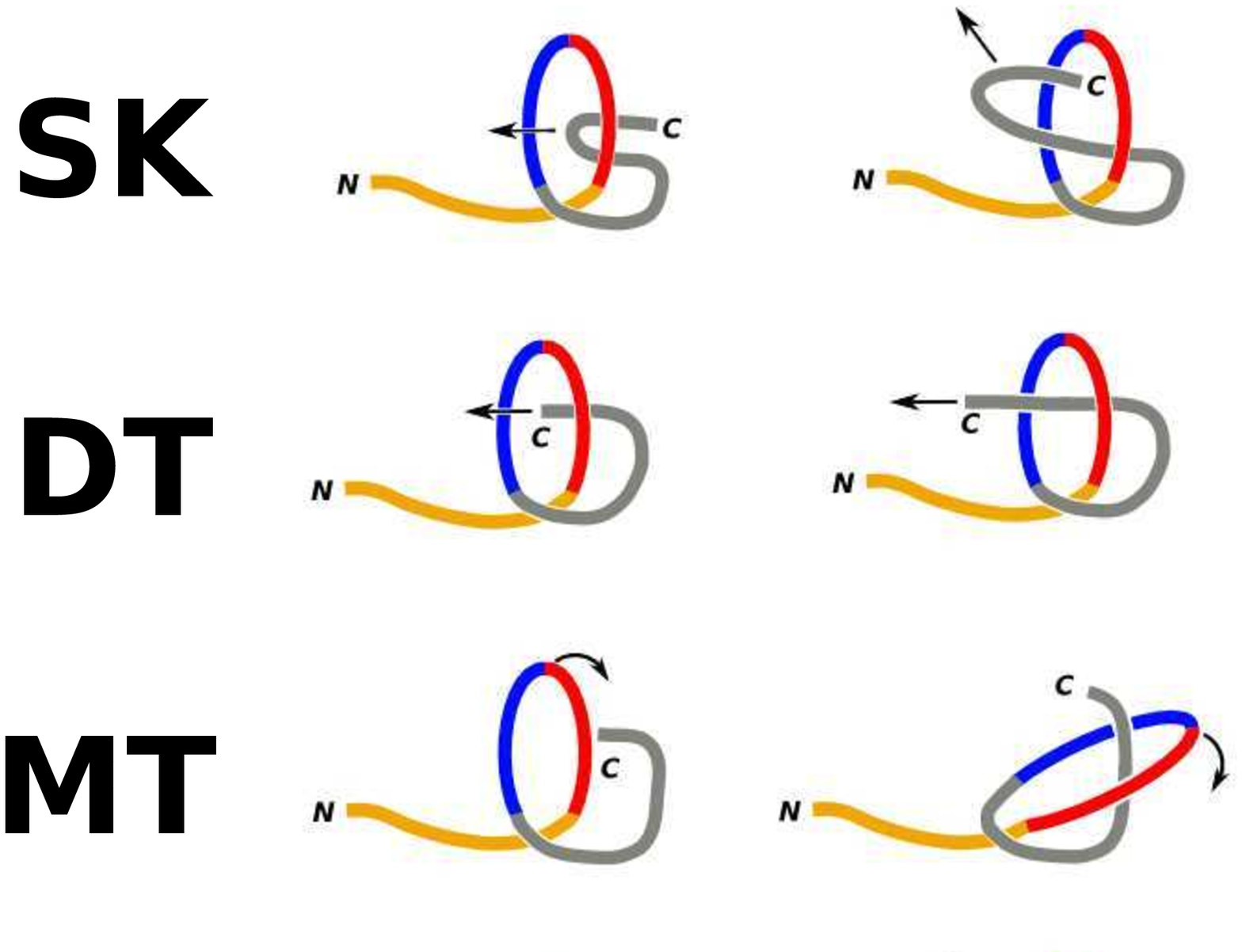}
\end{center}
  \caption{The folding mechanisms in 2EFV found in ref. \cite{Micheletti}.
SK denotes folding through slipknotting, DT -- through direct threading, and
MT -- through the mousetrap-like mechanism. Each process is illustrated by
showing two subsequent stages. SK involves sliding of a slipknot through the
knot-loop. In DT, the terminus threads through the knot-loop.
MT is similar to DT but the knot-loop 
makes the dominant movement instead of the terminal site.
The orange segment extends from the terminal N to site 16, the red segment --
from 17 to 53, the blue segment -- from 54 to 78, and the gray segment
is the remaining C-terminal piece.
}
  \label{mechsimple}
\end{figure}

In this paper, we focus on a protein with the native shallow trefoil knot:
MJ0366 from {\it Methanocaldococcus jannaschii} which is a thermophilic 
methanogenic archaea. 
This is the smallest of the known knotted proteins and we
shall refer to this protein by its
PDB structure code of 2EFV. Folding of 2EFV has been studied 
theoretically \cite{Micheletti,Noel} at a fixed temperature ($T$). The studies
involved all-atom simulations. Ref. \cite{Micheletti}
used a simplifying implicit solvent approach combined with a bias
implemented through the dominant reaction pathway method.
Out of 32 successful trajectories, 26 involved direct threading (DT), 
3 -- slipknotting (SK), and 2 -- mousetrapping (MT) as mechanisms of knotting.
These mechanisms are illustrated in Figure \ref{mechsimple}.
On the other hand, the simulations in ref. \cite{Noel} took a slipknot
conformation as the initial state of the system without exploring other
possible pathways: out of 15 40-$\mu$s-simulations, 5 resulted in correct folding.

Here, we use two variants of a structure-based coarse-grained model,
in which the protein is represented as a chain of beads
located at the $\alpha$-C positions,
and consider various $T$'s and much larger statistics of between 100
to 300 trajectories for each $T$. We show that:
a) 2EFV gets to the knotted native state much easier than the deeply
knotted proteins,
b) 2EFV should fold through a qualitatively richer family of pathways than considered in 
ref. \cite{Micheletti}, c) there is a range of optimal $T$'s for
successful knotting, d) the $T$-range corresponding to the fastest folding
is shifted downward relative to the $T$-range of the optimal knotting,
e) nascent conditions boost the peak success rate to fold 
in one of the variants of the model (in the other variant the off-ribosome
peak success rate is already 100\%), and f) nascent conditions reduce the
set of knotting pathways. 

All of the knotting mechanisms identified in ref. \cite{Micheletti} are
topologically single-stage processes in which just one knot-loop is formed.
We find that about 40\% of our successful trajectories indeed belong to this
class. However, the majority of the trajectories involve two stages
and two smaller knot-loops. Each of the stages makes use of variants
of the DT, and SK   
events, but we also identify one more - an 
"embracement" (EM). It is possible that the multiple-loop mechanisms of
knotting are also relevant for the homopolymer-like DNA, but have 
not been identified yet.

\section{Structure-based modeling}

The details of our approach are described 
in refs. \cite{JPCM, models, PLOS}. The model is Go-like \cite{Go0}
so that the length-related parameters in the potentials are derived from the native
structure. The molecular dynamics employed deals only with the $\alpha$-C atoms.
The bonded interactions are described by the harmonic potentials.
Non-bonded interactions, or contacts, are assigned to pairs
of amino acids by using the overlap criterion in which
the heavy atoms in the native conformation are represented by enlarged
van der Waals spheres \cite{JPCM,Tsai}: 
if at least two such spheres from different residues overlap we declare
existence of a native contact.
These contacts are described by
potentials with the minima
at the crystallographically determined distances. The potentials
are identical in depth, denoted as $\epsilon$. 
Non-native contacts are considered repulsive.

In order to test the robustness of our results,
we consider two variants of the model: C and A. 
In model C, the contact potentials are of the Lennard-Jones form and the
backbone stiffness is accounted for by the chirality potential \cite{JPCM}
which favors the native sense of the local backbone chirality. The value of $\epsilon$
has been calibrated by making comparisons to the experimental
data on stretching: approximately, $\epsilon$/{\AA} is 110 pN
(which also is close to the energy of the O-H-N hydrogen bond
of 1.65 kcal/mol).
In model A, the contact potentials are of the 10-12 kind and the
backbone stiffness is described by the more common bond and dihedral angles 
with the parameters specified in ref. \cite{Clementi}. 
Model C does not have the bond angle part of model A \cite{models}
and in model A, there are no $i,i+3$ contacts.
The simulations are done at various temperatures. For most unknotted proteins,
optimal folding takes place around
$T=0.3~ \epsilon/k_B$ in model C (see also ref. \cite{biophysical})
and around $T=0.6~ \epsilon/k_B$ in model A
($k_B$ is the Boltzmann constant; the stiffness parameters depend on $\epsilon$).
Both characteristic values of $T$ should correspond to a vicinity of the room $T$ 
in the respective models.

We use the Langevin thermostat with substantial damping.
The time unit of the simulations, $\tau$, is effectively of order 1 ns as 
the motion of the atoms is dominated by diffusion instead of being ballistic.
Folding is usually declared when all native contacts are established
for the first time (the distance between two $\alpha$-C in a contact
is smaller than the native distance multiplied by 1.5). 
For knotted proteins, however, this condition does 
not necessarily signify that the correct native knot has been formed.
The situation in which there is no knot but all contacts are established
is referred to as misfolding.

\section{Folding of 2EFV}

Protein 2EFV comprises 87 residues but the atomic coordinates of the
first five of them are not provided in the structure file.
In ref. \cite{Noel}, the authors extend the C-terminus 
(i.e. not where the residues are missing)
by a 5-piece helical segment to enhance the definition of the
starting slipknot  conformation. However, we find this procedure to
deteriorate folding properties so we show only the results obtained
without such an extension.
The secondary of 2EFV consists of 4 helices (23-32, 41-49, 62-71, 74-86)
2 3-10 helices (33-35, 72-73), and 2 $\beta$-strand (12-17, 54-59).
The knot ends in 2EFV are located at 11 and 73.  

\begin{figure}[ht]
\begin{center}
\includegraphics[scale=0.35]{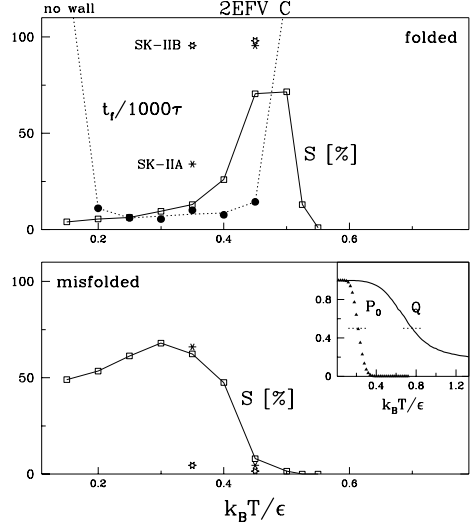}
\end{center}
  \caption{Properties of 2EFV in model C.
The inset in the lower panels shows $P_0$ and $Q$ as a function of $T$. The main
panels characterize the kinetic quantities: the upper panels are for correct folding
and the lower panels -- for misfolding. The open squares correspond to $S$ --
the success rate. The solid circles correspond to the median folding times.
These data points are obtained by starting from extended conformations.
The data points denoted by the hexagons and asterisks correspond the
the slipknotted conformations shown in Figure \ref{mechsimple}. 
}
  \label{chiral}
\end{figure}

Figure \ref{chiral} summarises the properties of 2EFV in model C.
The inset in the lower panel refers to the equilibrium quantities.
It shows the probability, $P_0$,
of all native contacts being \emph{simultaneously} established as a 
function of $T$. Similar to the lattice models of proteins \cite{Socci}
(see also an exact analysis \cite{Banavar}),
one may define $T_f$ as a temperature at which
$P_0$ crosses through $\frac{1}{2}$ -- it is 0.22 $\epsilon/k_B$ in this
model. The $T$ at which the fraction of the established native contacts, $Q$,
crosses through $\frac{1}{2}$ is higher, 0.75 $\epsilon/k_B$, as it
signifies the on-cooling onset of globular conformations. This
$T$ will be denoted as $T_Q$.

The top panel of Figure \ref{chiral} shows the
percentage-wise success, $S$, of reaching the properly knotted
folded conformation and the corresponding median folding time $t_f$
as a function of $T$. The median times have been determined only
within the subset of trajectories that resulted in folding. 
The starting conformations are nearly fully extended.
Folding is
seen to take place fast in the $T$-range between 0.2 and 0.55 $\epsilon/k_B$.
However, the majority of the trajectories result in knotting only
between 0.45 and 0.5 $\epsilon/k_B$ where $P_0$ is 0.  Thus the
peak success rates involve folding times that are longer than optimal.
At $T=0.3~\epsilon/k_B$,
$S$ is about 5\% which is still better than the peak success rates
reported for the deeply knotted proteins.
The bottom panel shows the $S$ corresponding to the misfolding events.
In the $T$-range corresponding to the optimal $t_f$'s most of the
trajectories result in misfolding. $S$ for correct folding and $S$
for misfolding do not add up to 100\% as a portion of the trajectories
does not establish all native contacts within a cutoff time of 1 000 000 $\tau$.

\begin{figure}[ht]
\begin{center}
\includegraphics[scale=0.35]{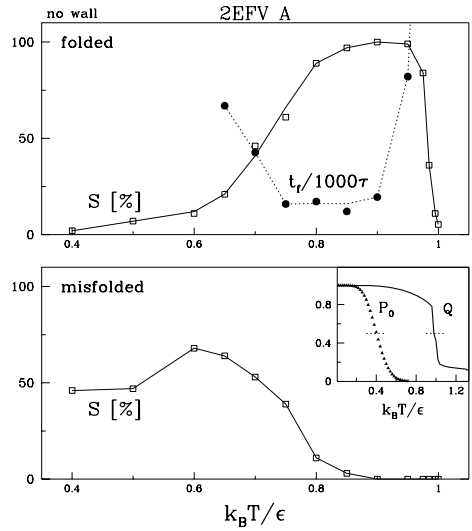}
\end{center}
  \caption{Similar to Figure \ref{chiral} but for model A.
}
  \label{angular}
\end{figure}

Figure \ref{angular} summarises the properties of 2EFV in model A.
The characteristic temperatures that relate to the kinetics move
upward by some 0.3 $\epsilon/k_B$ and $T_f$ shifts to 0.4 $\epsilon/k_B$
while $T_Q$ shifts to 1.0 $\epsilon/k_B$.
Even though the peak $S$ for folding is achieved just at the upper edge
of the kinetic optimality (at $T=0.9 \epsilon/k_B$), $S$ is larger than
50 \% in the whole range of the optimal kinetics (from 0.75 to 0.9 $\epsilon/k_B$.
The peak value of $S$ for misfolding (68 \%) is at $T=0.6 \epsilon/k_B$ and
it disappears at 0.9 $\epsilon/k_B$ in a gradual way.

Interestingly, we find that attachment of the
5-, 10-, 15-, and 20-residue N-terminal extensions
destroys the proper folding completely -- we would expect that
all-atom folding with the extensions should be even harder 
because of the many more degrees of freedom that need to cooperate.
However ref. \cite{Noel} states the opposite.
The extensions that we have considered are either random alanine segments
or a 5-residue helical extension (88-SER, 89-ALA, 90-ASN, 91-LEU, 92-LEU).
The misfolding events, on the other hand, are observed to be frequent.

\begin{figure}[ht]
\begin{center}
\includegraphics[scale=0.5]{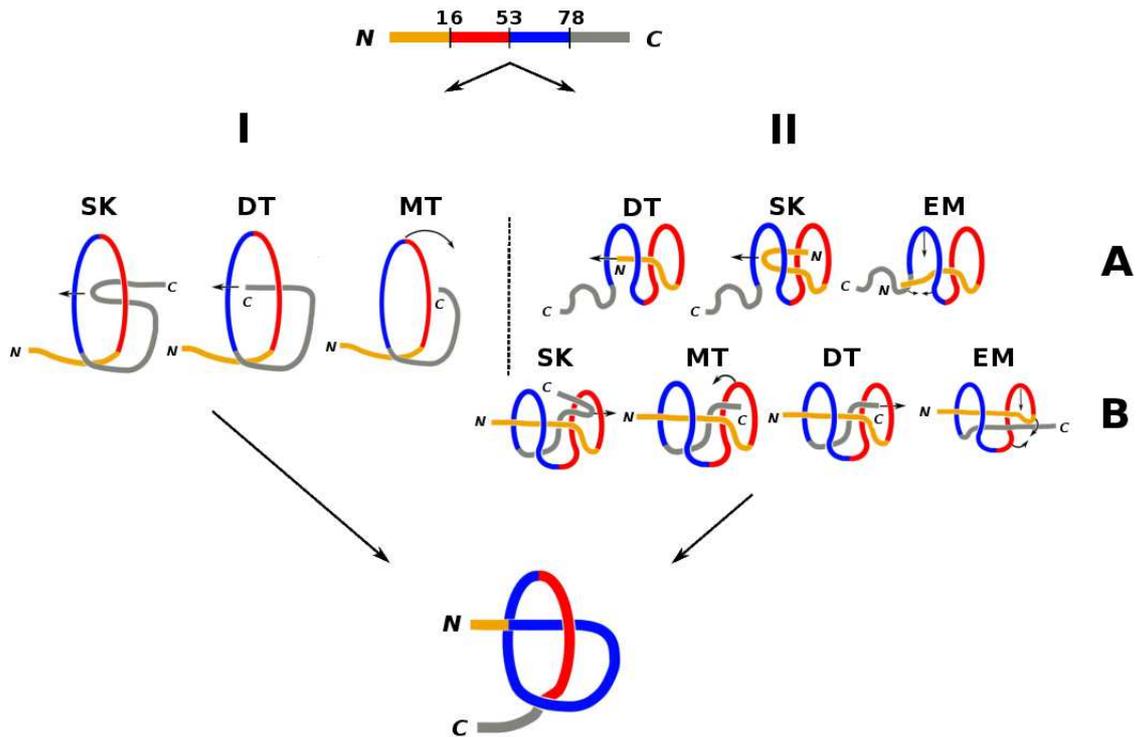}
\end{center}
  \caption{
Stages of folding and knotting process of protein 2EFV.
The top panel shows the extended conformation which is divided
into 4 sequential segments:
the N-terminal part is in yellow, C-teminal part is in gray, the first knot-loop
is in red, and the second knot loop in blue. The bottom panel
gives a schematic representation of the final native state.
Block I, on the left, shows the three single-loop mechanisms of knotting:
DT -- direct threading, SK -- slipknotting, MT -- mousetrapping.
Block II, on the right, shows two stages, A and B, of the two-loop
mechanisms of knotting. In addition to DT, SK, and MT, they also
involve EM --  embracement.
The smaller loops have radii between 8 and 10 {\AA}.
}
  \label{mechanisms}
\end{figure}

We observe that shallowly knotted proteins fold and knot
much easier than the deeply knotted ones -- but what are the mechanisms involved?
Figure \ref{mechanisms} shows that the proper pathways fall into two classes:
with a single knot-loop (the left hand side of the figure) 
formed from the segment between sites 16--78 or with two knot-loops
(the right hand side for the figure)  between sites 16--53 (shown in red) 
and 53--78 (in blue). The two classes arise in both models.
In model C, the two-loop mechanisms occur with the
$T$-averaged probability of 58\% and in model A -- 63\%. In the single-loop
class, the topological events involve the C-terminal parts. In the DT
mechanisms, the C-terminus threads through the knot-loop (45 \% in model C).
In the SK mechanism, the C-terminal slipknot slides through the knot-loop
(the mechanism invoked for the deeply knotted proteins; 45 \%).
In the MT mechanism, the knot-loop moves to envelop the C-terminus (10 \%).
A typical time scale in which the first  loop forms (also in the single-loop
mechanism) is around 4~000 $\tau$ in both models.
It takes much longer to form the second loop
-- typically 100 000 $\tau$ more. 
At high $T$'s (like $T=1.0~\epsilon/k_B$) in model A, it takes of order 500 000 $\tau$
to form a globular state followed a sudden single-loop knotting.

\begin{figure}[ht]
\begin{center}
\includegraphics[scale=0.6]{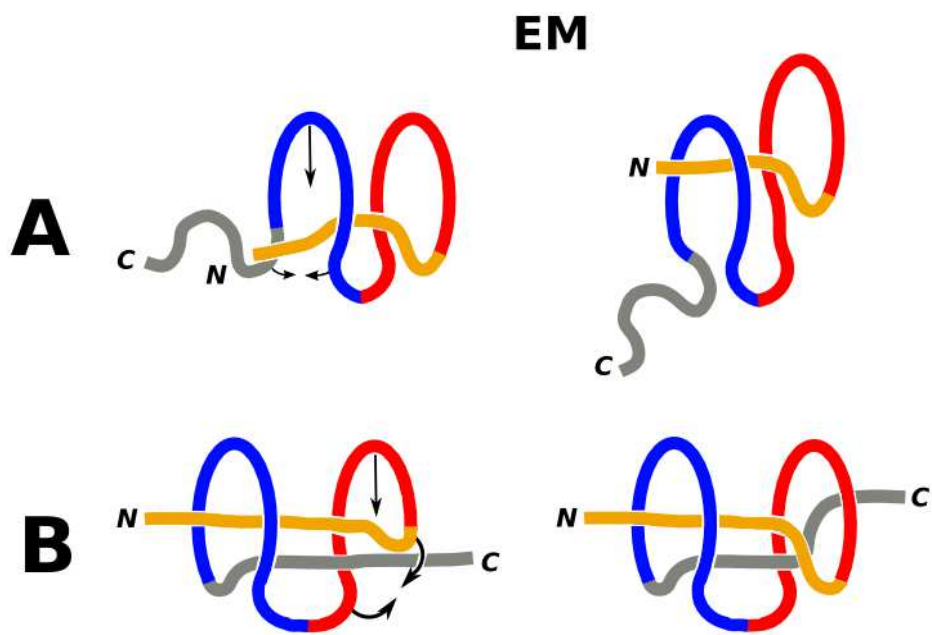}
\end{center}
  \caption{An example of a two-loop mechanisms of
folding combined with knotting. Here the embracement of the N-terminus
(the top panel) is followed by a similar event at the C-terminus
(the bottom panel).
The panels on the left show
the protein's conformation at the beginning of the knotting at the
particular stage. The right panels show the resulting state.
 }
  \label{embrace}
\end{figure}

In the two-loop class of pathways, two smaller knot-loops are formed and the events
involve both termini. Typically, the topological transformations
start at the N-terminus and, at stage A, 
split into three pathways.
They correspond to the mechanisms of DT (6 \%), SK (38 \%)  and EM (56 \%). 
The latter is one in which segment 53-78 "embraces" the mostly idle
N-terminal part and forms the knot-loop around it.
The next stage, denoted as B, knotting is completed by engaging
the C-terminal segment. Here, the pathways split into four mechanisms:
SK (44\%), MT (11\%), DT (39\%), and EM (6\%). 
There are some events in which stages A and B are interchanged.
An example of a situation in which one EM follows
another EM mechanism is shown in Figure \ref{embrace}.

In ref. \cite{Micheletti} the knotting mechanisms have been apparently
classified based on the events that shortly precede the appearance of knots
without identifying the one- and two-loop pathways.
If we cumulate the four IIB events with the I events and take the weighted
average between the two classes of pathways, we get about 44\% in SK,
41\% in DT, 11\% in MT, and 4\% in EM modes which does not agree with about
81\% in DT obtained in ref. \cite{Micheletti}. This could be due 
to either the role of the side groups, or the choice of just one
$T$ which may not be optimal, or the differences in the statistics.

We now consider a start from the slipknotted conformations 
without the extensions. If we start from SK - IIB, 
we get a 96\% success rate at $T=0.35 \epsilon/k_B$ and 98\% at $T=0.45~\epsilon/k_B$.
On the other hand, if we starts from SK - IIA then the success rates are 
correspondingly 34\% and 96\%. 
Finally, if we start from SK - I, similar to ref. \cite{Noel},
we get 97 \% at $T=0.45~\epsilon/k_B$. Our success rate is three times higher 
than in the all-atom simulations, which may reflect the Go-like 
character of our model and the lack of the water molecules.
However Noel {\it et al.} \cite{Noel} extend the protein at the C-terminus which
we find to deteriorate folding.

\section{Folding of 2EFV under nascent conditions}

\begin{figure}[ht]
\begin{center}
\includegraphics[scale=0.6]{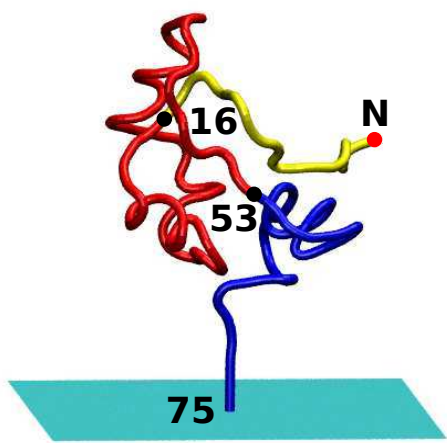}
\end{center}
  \caption{ 
Nascent protein 2EFV emerging from the model ribosome. At the stage shown, the segment
75 -- C is not yet born. The colors of the segments in the backbone 
approximate those of Figure \ref{mechanisms}.
 }
  \label{flower}
\end{figure}

The percentage-wise success, $S$, of reaching the properly knotted
folded conformation in model C increases substantially when simulating
the process in the co-translational way. 
In our previous model of on-ribosome folding \cite{nascent} we have
focused on the most essential aspects: the excluded volume
provided by the ribosome and the related reduction in the conformational
entropy are captured by representing the ribosome as an infinite
plate which spawns a protein residue by residue at one fixed  location.
We take the plate to generate a laterally uniform potential
of the form $\frac{3\sqrt{3}}{2}~\epsilon ~ (\frac{\sigma_0}{z})^9$,
where $z$ denotes the distance away from the plate and 
$\sigma_0 = 4\times 2^{-1/6}$.
The proteins are synthesized from the N terminus to the C terminus.
The time interval between the emergence of two successive $\alpha$-C atoms,
$t_w$, is taken as 5000 $\tau$ since larger values
lead to saturation in $S$. Once the backbone is formed fully, the
protein is released and then evolved up to a cutoff time of 1 000 000 $\tau$.
Our model is illustrated by Figure \ref{flower} which shows the protein
toward the end of the process in which it is being born.

\begin{figure}[ht]
\begin{center}
\includegraphics[scale=0.4]{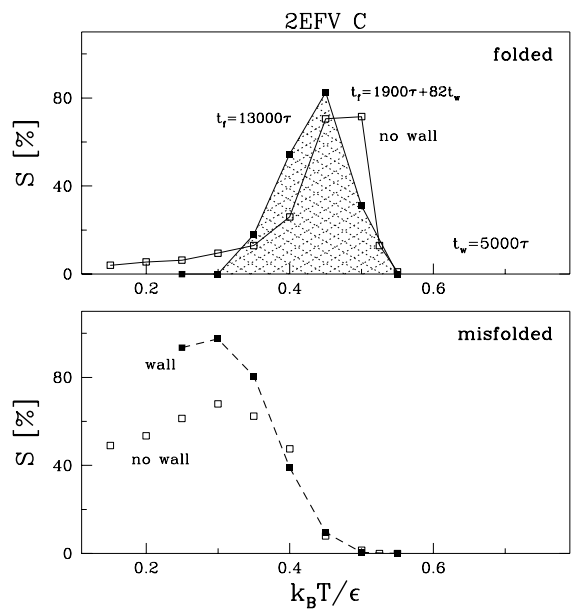}
\end{center}
  \caption{ The upper panel shows
the success rates for the on-ribosome (the solid squares)
and off-ribosome (the open squares) folding as a function of $T$.
The median folding times, $t_f$, are written for $T=0.4~\epsilon/k_B$
(13000~$\tau$ is for the off-ribosome folding).
The lower panel is similar but shows the
data for misfolding.
 }
  \label{sfol}
\end{figure}

Figure \ref{sfol} shows that the nascent conditions enhance
the probability of establishing the native contacts and the peak
success rate of knot formation. They also change the look of
the dependence of $S$ for knotting on $T$. 
For the deeply knotted protein the enhancement in $S$ made
a qualitative difference as it enabled folding \cite{nascent}. 
For the shallowly knotted 
2EFV, the kinetic improvement is minor -- by 5 percentage points from 
the peak value of $S$=76 \%. 
However, we observe a significant
shift in the occurrence of the topological events:
the single-loop pathways disappear and the role of the SK events,
so crucial for the deeply knotted proteins, gets diminished substantially.
Specifically, in step IIA shown 
in the right hand side of Figure \ref{mechanisms} the DT events disappear,
the SK events are reduced to 5\% and the EM mechanism is operational
in 95\% of the successful trajectories.
In stage IIB, MT and DT are enhanced (to 20 and 55\% respectively)
whereas SK gets diminished (25\%) and EM disappears.
The nascent conditions favor loops that first form at the N-terminus.

\section{Conclusions}

Our results for shallowly knotted 2EFV indicate easy and correct folding
with the topological optimality shifted upward in $T$ relative to the
kinetic optimality.
We have identified novel two-loop mechanisms of knotting and showed that
the topological mechanisms are much richer than discussed before.
Their variants may also be operational in homopolymers.
In our studies of YibK, we have found that extending the overlap-based
contact map by extra contacts identified by the CSU server \cite{Sobolev}
was vital for the success of on-ribosome folding. However, for 2EFV
adding the extra contacts is found not to affect any of the results
discussed here. On-ribosome folding is more efficient than 
off-ribosome and is also more selective in its mechanisms.
It would be interesting to consider a
realistic variant of co-translational folding --
one that takes into account confinement. This is because the
ribosome spawns proteins into molecularly sculpted chambers \cite{Elcock},
which parallels the physics
of encapsulation within GroEL-GroES chaperonins considered by Lim and Jackson
\cite{LimJackson} in the context of deeply knotted proteins.

{\bf Acknowledgments}
We appreciate useful correspondence with C. Micheletti and
J. Trylska.
This work has been supported by the National Science Center in Poland
under the aegis of the EU Joint Programme
in Neurodegenerative Diseases (JPND).
The local computer resources were financed by the European Regional Development Fund 
under the Operational Programme Innovative Economy NanoFun POIG.02.02.00-00-025/09.

\clearpage

\end{document}